\documentclass[preprint,aps,showpacs,nofootinbib,preprintnumbers,amsmath,amssymb,superscriptaddress]{revtex4-1}
\usepackage{amssymb}
\usepackage{epsfig}
\usepackage{latexsym}
\usepackage{graphicx}
\usepackage{subfigure}
\usepackage{dcolumn}
\usepackage{bm}
\usepackage[usenames ,dvipsnames]{xcolor}
\usepackage[utf8]{inputenc}
\usepackage{slashed}
\usepackage{cleveref}
\newcommand{\sme}{\sigma_{\text{ex}}}
\newcommand{\pc}{\varphi_{c}}

\begin{document}

\title{Scalar Gravitational Wave Signals from Core Collapse in  Massive Scalar-Tensor Gravity with Triple-Scalar Interactions}

\author{
  Da~Huang\footnote{dahuang@bao.ac.cn}
}
\affiliation{National Astronomical Observatories, Chinese Academy of Sciences, Beijing, 100012, China}
\affiliation{School of Fundamental Physics and Mathematical Sciences, Hanzhou Institute for Advanced Study, UCAS, Hanzhou 310024, China}
\affiliation{International Centre for Theoretical Physics Asia-Pacific, Beijing/Hanzhou, China}

\author{
Chao-Qiang~Geng\footnote{geng@phys.nthu.edu.tw}
}
\affiliation{School of Fundamental Physics and Mathematical Sciences, Hanzhou Institute for Advanced Study, UCAS, Hanzhou 310024, China}
\affiliation{International Centre for Theoretical Physics Asia-Pacific, Beijing/Hanzhou, China}
\affiliation{Department of Physics, National Tsing Hua University, Hsinchu 300, Taiwan}
\affiliation{Synergetic Innovation Center for Quantum Effects and Applications (SICQEA), 
Hunan Normal University, Changsha 410081, China}

\author{Hao-Jui Kuan \footnote{guanhauwzray@gmail.com}}
\affiliation{Department of Physics, National Tsing Hua University, Hsinchu 300, Taiwan}
\affiliation{Theoretical Astrophysics, IAAT, University of Tubingen, Tubingen, Germany}

\date{\today}
\begin{abstract}
The spontaneous scalarization during the stellar core collapse in the massive scalar-tensor theories of gravity introduces extra polarizations (on top of the plus and cross modes) in gravitational waves, whose amplitudes are determined by several model parameters. Observations of such scalarization-induced gravitational waveforms therefore offer valuable probes into these theories of gravity. Considering a triple-scalar interactions in such theories, we find that the self-coupling effects suppress the magnitude of the scalarization and thus reduce the amplitude of the associated gravitational wave signals. In addition, the self-interacting effects in the gravitational waveform are shown to be negligible due to the dispersion throughout the astrophysically distant propagation. As a consequence, the gravitational waves observed on the Earth feature the characteristic inverse-chirp pattern. Although not with the on-going ground-based detectors, we illustrate that the scalarization-induced gravitational waves may be detectable at a signal-to-noise ratio level of ${\cal O}(100)$ with future detectors, such as Einstein Telescope and Cosmic Explorer.

\end{abstract}

\maketitle

\section{Introduction}
\label{s1}
Even though general relativity (GR) has so far withstood all experimental tests~\cite{Psaltis:2008bb,Berti:2015itd, Will:2014kxa, TheLIGOScientific:2016src}, many theoretical considerations signal the need of alternative theories of gravity beyond the GR, e.g., the non-renormalizability of GR~\cite{Burgess:2003jk}, and numerous astrophysical and cosmological observations~\cite{Spergel:2015noa} (see Ref.~\cite{Berti:2015itd} for a recent review).
 An abundance of modified theories of gravity have been constructed to extend GR to overcome the aforementioned obstacles. Among them, the most widely explored one may be the scalar-tensor (ST) theories of gravity~\cite{Jordan:1959eg, Fierz:1956zz, Brans:1961sx} in which the gravitational interaction is mediated not only by the metric tensor but also by an additional scalar field. 

Due to its simplicity and well-posedness~\cite{Salgado:2005hx, Salgado:2008xh}, ST theories have been subjected to a wide range of experimental tests~\cite{Damour:1992we,Chiba:1997ms,Fujii:2003pa,Faraoni:2004pi, Sotiriou:2015lxa}, while most are performed in the weak-field regime (mainly in the solar system)~\cite{Shapiro:1999fn,Williams:2005rv,Shapiro:2004zz,Bertotti:2003rm}. Nonetheless, with appropriate ST parameters and strong enough spacetime curvature (see, e.g., \cite{Ventagli20}), strongly-scalarized neutron star (NS) solutions are energetically favoured over the ordinary weakly-scalarized ones \cite{Salgado98}. The ST gravity therefore allows a nonperturbative strong-field phenomenon known as the {\it spontaneous scalarization}~\cite{Damour:1993hw,Damour:1996ke} in NSs. Such phenomenon induces a non-vanishing dipole charge of the scalar field, which is (stringently) constrained by observations such as the Cassini spacecraft~\cite{Bertotti:2003rm} and binary pulsar timings~\cite{Wex:2014nva,Freire:2012mg,Antoniadis:2013pzd}. However, in massive ST theories, the introduction of the scalar mass $\mu \gtrsim 10^{-15}$~eV effectively weakens these experimental constraints on the parameters~\cite{Ramazanoglu:2016kul,Morisaki:2017nit} so that the theory leaves more room for parameters that can lead to strongly-scalarized NS solutions. In addition, the recent detections of gravitational waves (GWs) have opened a novel window to limit the ST gravity, which has already set severe constraints on the possible models~\cite{TheLIGOScientific:2016src,Yunes:2016jcc,Abbott:2018lct,LIGOScientific:2019fpa}. A further constraint on the ST gravity arises from the equivalence between the Jordan and Einstein frames~\cite{Geng:2020slq,Geng:2020ftu}.

As the evolutionary endpoint of a star, the core collapse to form a NS or a black hole (BH) constitutes another testbed for the strong-field dynamics of the ST gravities~\cite{Berti:2015itd,Ramazanoglu:2016kul,Andreou:2019ikc}. In particular, if a scalarized proto-NS forms during the collapse, the sharp transition from the weakly-scalarized star configuration into the strongly-scalarized state generates scalar- or monopole-polarized GWs, which is a critical feature of the ST gravity that is absent in GR. Early studies on the spontaneous scalarization and scalar GW production were concentrated on the massless ST gravity~\cite{Novak:1997hw,Novak:1998rk,Novak:1999jg,Gerosa:2016fri}. In order to evade the strong experimental constraints listed above, Refs.~\cite{Sperhake:2017itk,Rosca-Mead:2020ehn} initiated a line of studies on the core collapse in the massive ST theories. In Refs.~\cite{Cheong:2018gzn,Rosca-Mead:2019seq}, the theory was further extended to the case with scalar self-interactions which generically suppressed the degree of the scalarization and the amplitude of the scalar GW radiation. 

However, previous studies only focused on the effects of the even-power scalar self-interactions~\cite{Cheong:2018gzn,Rosca-Mead:2019seq} defined in the Einstein frame. Odd-power scalar self-interactions have not been explored yet. Given the importance of the stellar collapse in testing ST gravity by, e.g., scalarization-induced GW signals that are potentially measurable with the ground-based GW detectors, any ST model warrants further investigations. In the present paper, we study the core collapse and the monopole GW radiations in the massive ST theories, including the simplest odd-power scalar self-interaction in the Einstein frame, viz.~the triple-scalar coupling effects. In order to ensure the stability of the theory, we take the absolute value of the triple-scalar interaction in the scalar potential. The influence of this scalar interaction on the spontaneous scalarization and the scalar GW production during the core collapse is explored by considering two particular collapse progenitors, which have been treated in the massless \cite{Gerosa:2016fri} and massive \cite{Sperhake:2017itk,Geng:2020slq} ST theories. One of them leads to a NS remnant and the other settles to a BH. We also consider the impact of the triple-scalar coupling on the astrophysically long-distance dispersive propagation of the scalar GWs in our galaxy, which generically leads to the inverse-chirp feature of the GW signals detected on the Earth. Finally, we assess the detectability of the possible scalar GW signals in the on-going and up-coming ground-based GW detectors by computing the corresponding signal-to-noise ratios (SNRs). 

The paper is organized as follows. In Sec.~\ref{formalism}, we present a brief introduction to the self-interacting massive ST theories and specify the triple-scalar interaction that is considered. The equation of state obeyed by the nuclear matter during the stellar core collapse is introduced as well. In Sec.~\ref{ModelResult}, the setup of our simulations is detailed. We also show the numerical results of the core collapses with a specific attention paid to the self-interaction effects on the scalar dynamics and the associated GW waveforms. Sec.~\ref{GW_PSD} is devoted to studying the influence of the triple-scalar coupling on the large-distance propagation of the scalar GWs and the detectability of these GW signals. Finally, we offer our conclusions and some further discussions in Sec.~\ref{Conclusion}.    

\section{Formalism}\label{formalism}
In the scalar-tensor theories of gravity, the gravitational fields are augmented by an extra scalar field $\phi(x)$ in addition to the spacetime metric tensor $g_{\mu\nu}(x)$. In this work, we are interested in the ST gravity first proposed by Bergmann~\cite{Bergmann:1968ve} and Wagoner~\cite{Wagoner:1970vr}, in which (i) the single scalar field non-minimally couples to the metric tensor; (ii) the theory is diffeomorphism invariant; (iii) the variation of the action gives rise to second order field equations; and (iv) the weak equivalence principle is satisfied [in the Jordan (physical) frame, see below]. In the literature, the most general action of this class of theories can be written as~\cite{Berti:2015itd}
\begin{eqnarray}\label{eq:Sjord}
S_J = \int d^4 x \frac{\sqrt{-g}}{16\pi} \left(F(\phi) R - \frac{\omega(\phi)}{\phi} g^{\mu\nu} \partial_\mu \phi \partial_\nu \phi - U(\phi)\right)+S_m[\psi_m, g_{\mu\nu}]\,,
\end{eqnarray}
in natural units $G=c=1$, where $F(\phi)$, $\omega(\phi)$ and $U(\phi)$ are regular functions of the scalar field $\phi$, and $S_m$ represents the action of ordinary matter fields that are collectively denoted by $\psi_m$. Note that matter fields couple to the gravity sector only through the (physical) metric tensor $g_{\mu\nu}$ without any dependence on $\phi$, so that the weak equivalence principle holds.

One can also formulate the same theory in the so-called Einstein's frame, which relates to the Jordan frame via a Weyl transformation,
\begin{align}
	\bar{g}_{\mu\nu} \equiv g_{\mu\nu}/F(\varphi),	
\end{align}
and a redefinition of the scalar field,
\begin{align}
	\frac{d\varphi}{d\phi} := \pm \sqrt{\frac{3(F_{,\phi})^2}{4 F^2}+\frac{\omega}{2\phi F}}.
\end{align}
The action Eq.~\eqref{eq:Sjord} reads
\begin{eqnarray}\label{ActionEinstein}
S_E = \int d^4 x \frac{\sqrt{-\bar{g}}}{16\pi}\Big(\bar{R} -2 \bar{g}^{\mu\nu} \partial_\mu \varphi \partial_\nu \varphi - 4V(\varphi)\Big)+ S_m [\psi_m, \bar{g}_{\mu\nu}/F(\varphi)]\,,
\end{eqnarray}
in the Einstein frame, where the scalar potential $V(\varphi)$ is related to the one in the Jordan frame through~\cite{Geng:2020slq,Geng:2020ftu}
\begin{eqnarray}
V(\varphi):= U(\phi)/(4F^2)\,.
\end{eqnarray}
One can see that the ST theory in the Einstein frame is specified by the scalar potential $V(\varphi)$ and the conformal factor $F(\varphi)$. The latter controls the coupling between the scalar field $\varphi$ and ordinary matters.
In the present article, we adopt the standard parametrization,
\begin{eqnarray}\label{ConformalFactor}
\ln F(\varphi) \equiv -2\alpha_0 (\varphi-\varphi_0) - \beta_0 (\varphi-\varphi_0)^2\,,
\end{eqnarray}
of the conformal factor ~\cite{Damour:1992we, Damour:1996ke,Chiba:1997ms}, where $\alpha_0$ and $\beta_0$ are two free dimensionless parameters determining the modifications of gravity in the first post-Newtonian order, and $\varphi_0$ is the asymptotic value of the scalar field $\varphi$ at spatial infinity. 

Massive scalar fields with a triple-scalar self-interaction have the scalar potential given by
\begin{eqnarray}\label{potential}
V(\varphi) = \frac{\mu^2}{2\hbar^2}(\varphi^2+ \lambda |\varphi|^3 )\,,
\end{eqnarray}
where $\lambda$ is assumed to be positive and the absolute value is taken in the scalar self-interaction term in order to guarantee the semi-positivity and stability of the potential. Thus, the asymptotic value of $\varphi$ should naturally be $\varphi_0 = 0$, which is the absolute minimum of this potential. Moreover, the scalar field mass is fixed to be $\mu = 10^{-14}$~eV, which not only alleviates the strong constraints from binary pulsars and weak field tests of GR~\cite{Ramazanoglu:2016kul,Morisaki:2017nit}, but also allows the propagating monopole GW signals to be detectable~\cite{Sperhake:2017itk,Rosca-Mead:2020ehn} in the LIGO/Virgo sensitivity window. For the later convenience, we also define a characteristic frequency
\begin{align}\label{eq:comp}
	f_* = \frac{\omega_*}{2\pi} \equiv \mu/(2\pi\hbar) \simeq 2.42 \text{ Hz},
\end{align}
which is associated with the scalar mass.  

\subsection{Spherically Collapse}
The stellar core collapse into a compact object such as NS and BH marks the endpoint of a massive star with the zero-age main sequence (ZAMS) mass in the range of $10 M_\odot \lesssim M_{\rm ZAMS} \lesssim 130 M_\odot$~\cite{OConnor:2010moj, Clausen:2014wia, Sukhbold:2015wba} with $M_\odot$ denoting the solar mass. At the end of the nuclear burning phase of a star, the thermal pressure from the nuclear matter can no longer balance the huge gravitational attracting force~\cite{Bethe:1990mw}, leading to a sudden radial compression of the matter. The collapse increases the central (baryon) mass density up to the nuclear density $\rho_{\rm nuc} \simeq 2\times 10^{14}~{\rm g\cdot cm}^{-3}$~\cite{Dimmelmeier:2002bm} beyond which the compression is halted and the inner core bounces due to the repulsive force stemmed from the stiffening of the nuclear matter. The core bounce launches a hydrodynamic shockwave outwards and a core-collapse supernova explostion (CCSNe) may then be generated\footnote{After emanating from the inner core, the out-going shock will then be stalled by the inflowing material. If the shockwave successfully revives by some mechanisms, such as the standing accretion shock instability \cite{Burrows06,Harada17}, and the neutrino heating \cite{Janka01,Pejcha12}, the explosion will be instigated and lead to a CCSNe. The mechanism that accounts for the revival is still an on-going problem (see, e.g., \cite{Couch17} for a recent review and the references therein).}. 

We assume a spherically symmetric anatz for which the line element is given by~\cite{Bardeen:1982de}
\begin{eqnarray}
ds^2 = -F \alpha^2 dt^2 +F X^2 dr^2 +r^2(d\theta^2 + \sin^2\theta d\phi^2)\,,
\end{eqnarray}
where $X$ and $\alpha$ are functions of the coordinates $r$ and $t$. The stellar matter is modelled as the a perfect fluid whose energy-momentum tensor reads~\cite{OConnor:2009iuz}
\begin{eqnarray}
T_{\mu \nu } = (\rho+\rho \epsilon +P) u_\mu u_\nu + P g_{\mu\nu}\,,
\end{eqnarray}
in the Jordan frame, where $\rho$, $P$, $\epsilon$, and $u^\mu$ represent the baryon density, pressure, internal energy and the four-velocity of the nuclear matter, respectively. In order to account for the stiffness of the nuclear matter and thermal effects experienced by the shock, we would like to follow Refs.~\cite{Gerosa:2016fri, Sperhake:2017itk,Rosca-Mead:2020ehn} to apply the hybrid equation of state (EOS)~\cite{Janka1, Zwerger:1997sq,Dimmelmeier:2002bm}
\begin{eqnarray}
P = P_c + P_{\rm th}\,, \quad \quad \epsilon= \epsilon_c + \epsilon_{\rm th}\,,
\end{eqnarray}
which comprise the cold parts ($P_{\text{c}},\epsilon_{\text{c}}$) and the thermal parts ($P_{\text{th}},\epsilon_{\text{th}}$). The cold parts are given by
\begin{eqnarray}
P_c &=& K_1 \rho^{\Gamma_1} \,, \quad \epsilon_c = \frac{K_1}{\Gamma_1 -1} \rho^{\Gamma_1 -1}\,,\quad {\rm as}\,\,\rho\leq \rho_{\rm nuc}\,,\nonumber\\
P_c &=& K_2 \rho^{\Gamma_2}\,, \quad \epsilon_c = \frac{K_2}{\Gamma_2 -1} \rho^{\Gamma_2 -1}+E_3\,,\quad {\rm as}\,\, \rho > \rho_{\rm nuc}\,,
\end{eqnarray}
with $K_1 = 4.9345 \times 10^{14}$~[cgs], while $K_2$ and $E_3$ are determined by the continuity of the pressure and energy density at $\rho=\rho_{\text{nuc}}$. The thermal parts that describe a mixture of non-relativistic and relativistic gases generate the pressure as $P_{\rm th} = (\Gamma_{\rm th}-1)\rho \epsilon_{\rm th}$. The nuclear matter EOS is determined by the three parameters $(\Gamma_1, \Gamma_2, \Gamma_{\rm th})$, which are fixed to be $(1.3, 2.5, 1.35)$ by following Ref.~\cite{Geng:2020slq}.

Our numerical simulations for stellar core collapses are based on the open-source code {\tt GR1D}~\cite{OConnor:2009iuz}, which was originally developed to model the spherically symmetric hydrodymics in GR by exploying the high-resolution shock capturing scheme~\cite{Font:1998hf,Font:2008fka}. The code was extended to include a massless scalar field in Ref.~\cite{Gerosa:2016fri} and a massive one in Refs.~\cite{Sperhake:2017itk,Rosca-Mead:2020ehn}, and was further implemented with an even-power scalar potential in Refs.~\cite{Cheong:2018gzn,Rosca-Mead:2019seq}. Here we adopt all dynamical equations, grid types and boundary conditions in our simulations identical to those in Ref.~\cite{Sperhake:2017itk}. 

 
\section{Simulations and Results}\label{ModelResult}
We focus on two specific progenitors of supernovae (pre-SN) with primordial metallicities of $M_{\rm ZAMS} = 12 M_\odot$ (denoted by WH12) and $M_{\rm ZAMS} = 40 M_{\odot}$ (denoted by WH40) from the catalog of realistic non-rotating pre-SN models of Woosley and Heger (WH)~\cite{Woosley:2007as},
respectively. In massive ST gravity~\cite{Gerosa:2016fri}, as well as in GR~\cite{Kuan21}, it has been shown that WH12 will collapse to form a NS remnant, while WH40 will eventually result in a BH after a stage of proto-NS (though see later).

The yet undermined ST parameters $\alpha_0$ and $\beta_0$ in the conformal factor of Eq.~(\ref{ConformalFactor}) are chosen to allow strong scalarization in proto-NSs so as to generate significant scalar GW signals in our simulations. Their specific values will be given along with the detailed discussion of these two cases in Section \ref{sec.III.A}. Note that the initial value of the scalar field at every grid point is taken to be zero. 

\subsection{Numerical Results on Core Collapse to Neutron Stars} \label{sec.III.A}

For the particular progenitor WH12, the ST parameters are fixed to be $\alpha_0 = 10^{-2}$ and $\beta_0 = -20$, which trigger the  strong scalarization in the remnant NS as shown in Refs.~\cite{Sperhake:2017itk,Cheong:2018gzn,Geng:2020slq}. Our main interest here is to investigate how the triple-scalar interaction in Eq.~(\ref{potential}) affects the scalar field dynamics during the core collapse and its subsequently produced GW signals. 

\begin{figure}[!htb]
	\centering
	\hspace{-5mm}
	\includegraphics[scale=0.4]{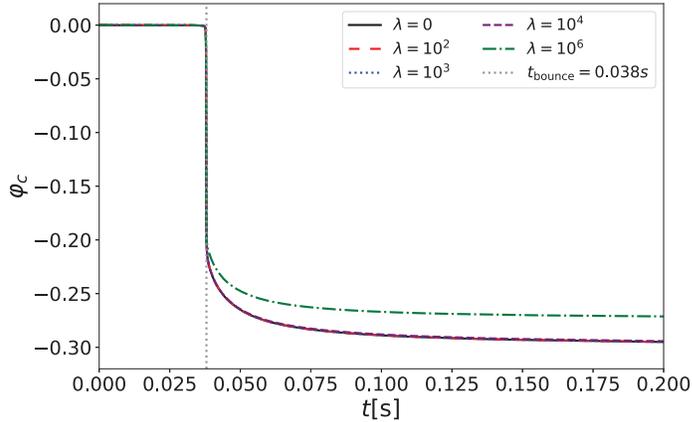}	
	\caption{Evolutions of the central value of the scalar field $\varphi_c$ during the core-collapse into a NS for the model WH12 with  the self-interaction couplings
	of  $\lambda =$0, $10^2$, $10^3$, $10^4$ and $10^6$, respectively, where the vertical dashed line denotes the time of the stellar core bounce.} \label{central_NS}
\end{figure}	

In Fig.~\ref{central_NS}, we show the scalar field value at the stellar center as a function of time, {\it i.e.}, $\varphi_c (t) \equiv \varphi(t,r=0)$. We can see that 
the strong spontaneous scalarization occurs at the time of the core bounce, $t\simeq 0.038$~s, for all triple-scalar coupling strength $\lambda$ considered. The departure of the scalar dynamics around the star center from the non-interacting ST gravity ($\lambda = 0$) is found to be inconsiderable for a moderate coupling strength $\lambda \lesssim 10^4$. However, when $\lambda$ approaches or exceeds $\sim 10^6$, we can observe the significant suppression in the magnitude of the scalarization (the green dashed line in Fig.~\ref{central_NS}).

\begin{figure}[!htb]
	\centering
	\hspace{-5mm}
	\includegraphics[width = 0.5 \linewidth]{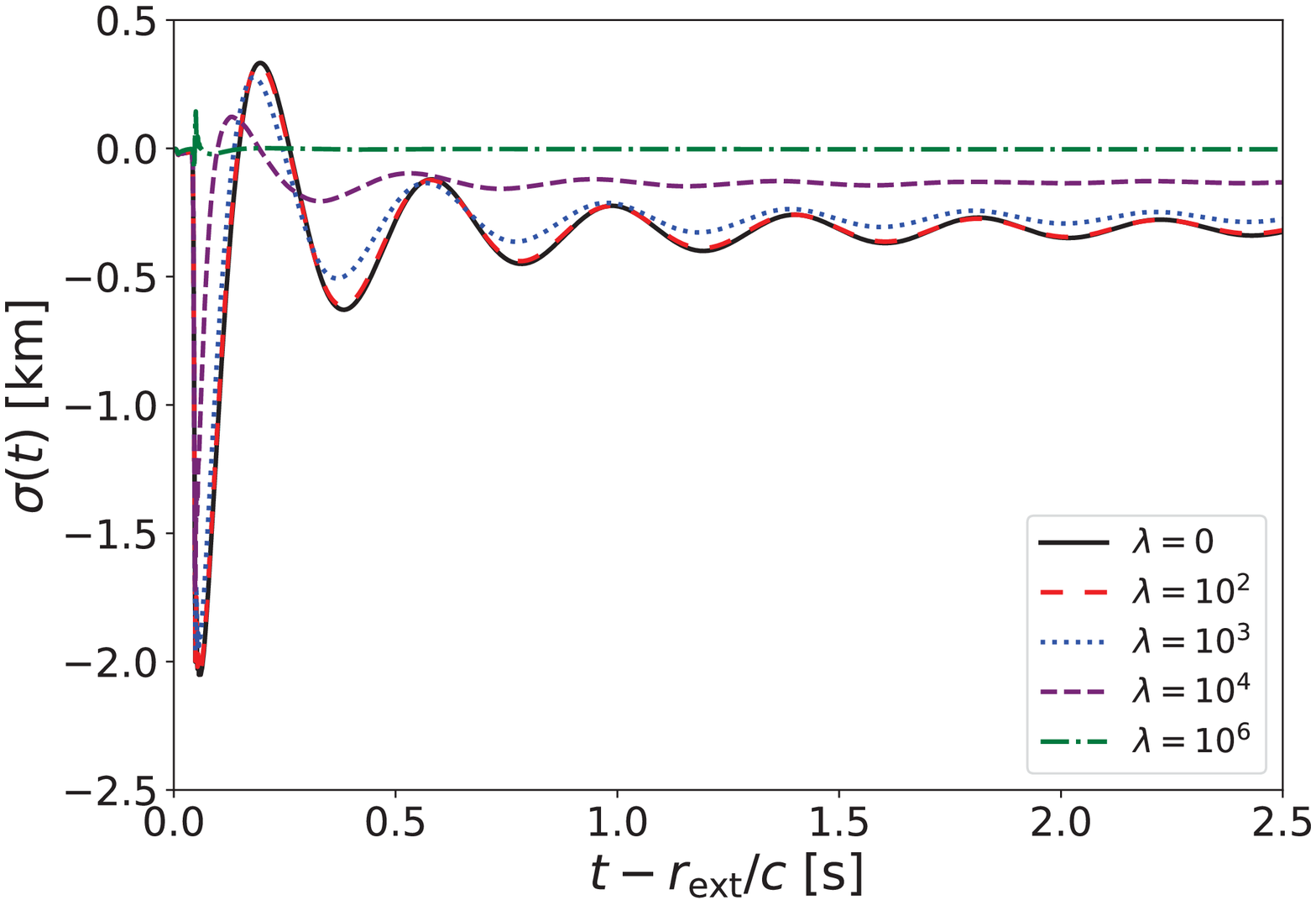}	
	\includegraphics[width=0.5\linewidth]{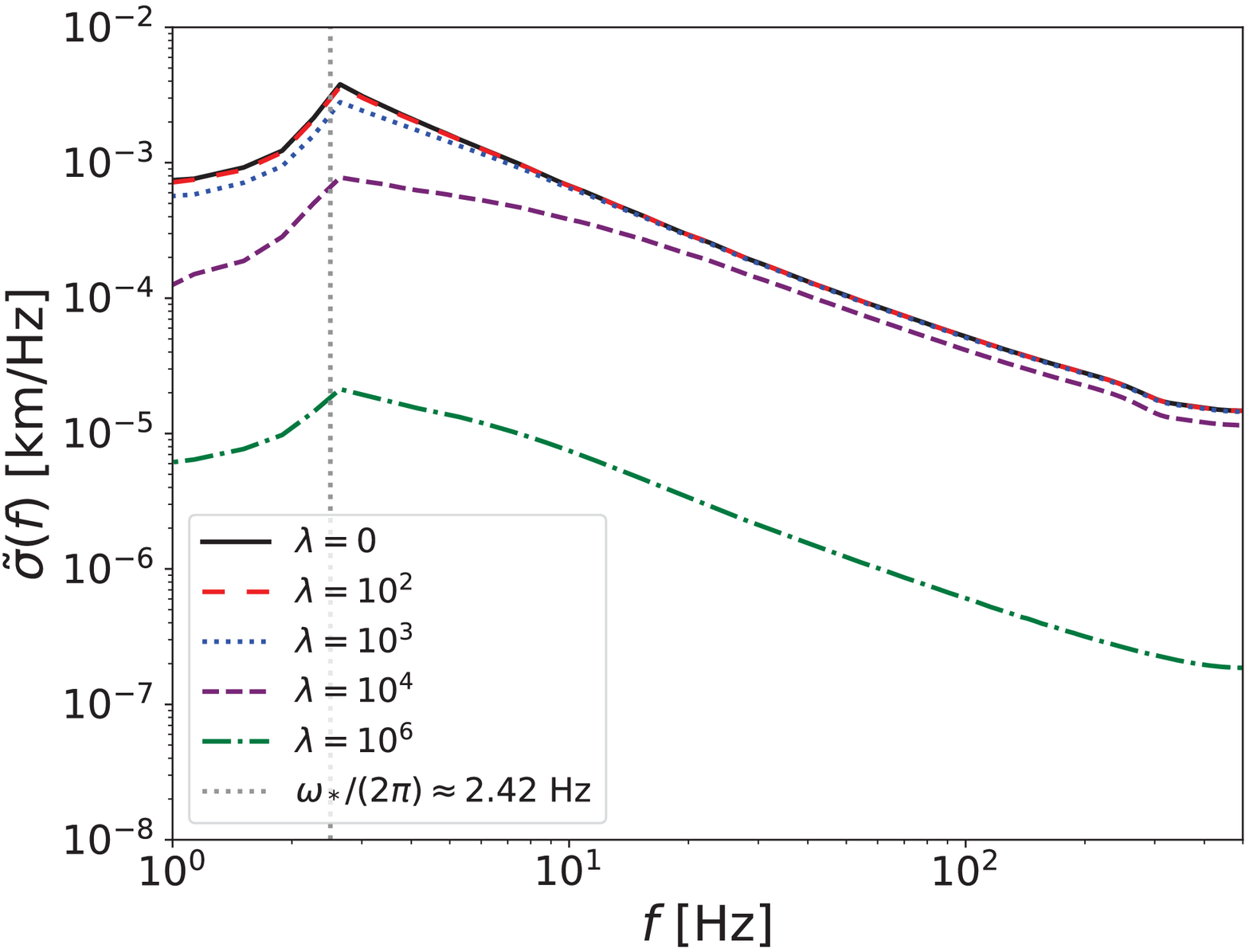}
	\caption{Scalar GW signals for WH12 extracted at the radius $r_{\rm ex} = 5\times 10^{4}$~km for  the self-interaction couplings
	of  $\lambda =$0, $10^2$, $10^3$, $10^4$ and $10^6$, respectively, where the vertical line on the right plot denotes the characteristic frequency $f_*= 2.42$~Hz.} \label{benchmark_NS}
\end{figure}	

The suppression of the spontaneous scalarization is more evident in the rescaled field 
\begin{align}
	\sigma \equiv r\varphi.
\end{align}
According to its definition, the leading-order term of $\sigma$ reduces to the scalar charge $\chi$ in the massless ST gravity \cite{Gerosa:2016fri}, which is defined as the coefficient of the $1/r$ term in the asymptotic behavior of the scalar field at spatial infinity, i.e.,
\begin{align}
	\varphi=\frac{\chi}{r} + O(r^{-2}).
\end{align}
Extracting $\sigma$ at a fixed distance of $r_{\rm ex}= 5\times 10^4\text{ km}$, in Fig.~\ref{benchmark_NS} we plot $\sme \equiv \sigma(r_{\rm ex})$ in the time domain (left panel), and the associated Fourier transformed signals $\tilde{\sigma}_{\text{ex}}(f)$ \cite{Ivezic1} in the frequency domain (right panel). For the self-coupling up to $\lambda \sim 10^2$, the difference in $\sme$ from the non-self-interacting ST gravity is negligible. However, when $\lambda$ approaches $\sim 10^3$, we begin to see a mild suppression and a small phase shift. The suppression and the phase shift are enlarged with increasing $\lambda$ as a consequence of the frequency-related suppression caused by the scalar self-interaction (similar to the fashion of  Ref.~\cite{Cheong:2018gzn}). Finally, when $\lambda\gtrsim10^6$, the scalarizaion is reduced from the case without the self-interaction by more than two orders.

\begin{figure}[!htb]
	\centering
	\hspace{-5mm}
	\includegraphics[width = 0.5 \linewidth]{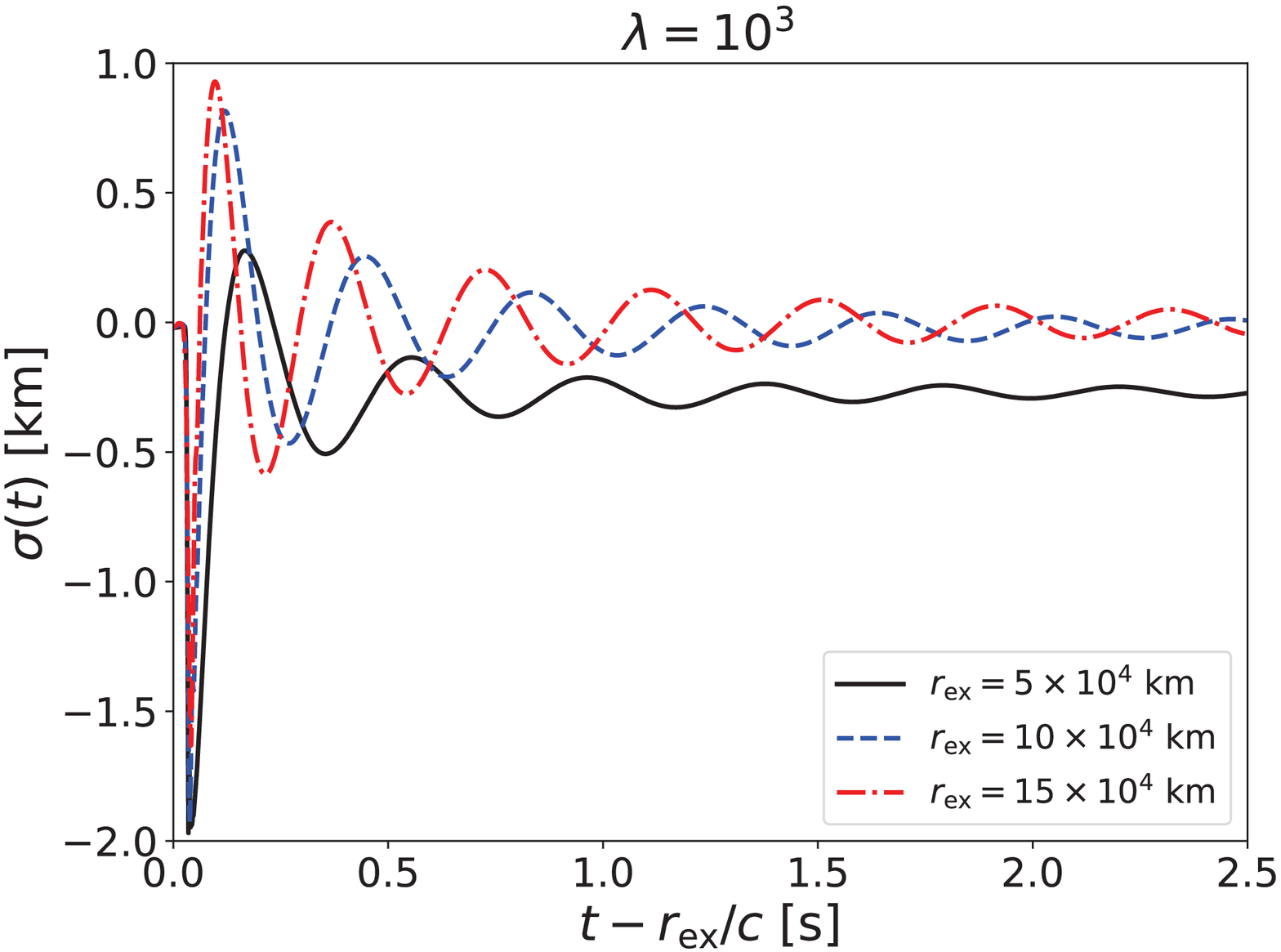}	
	\includegraphics[width=0.5\linewidth]{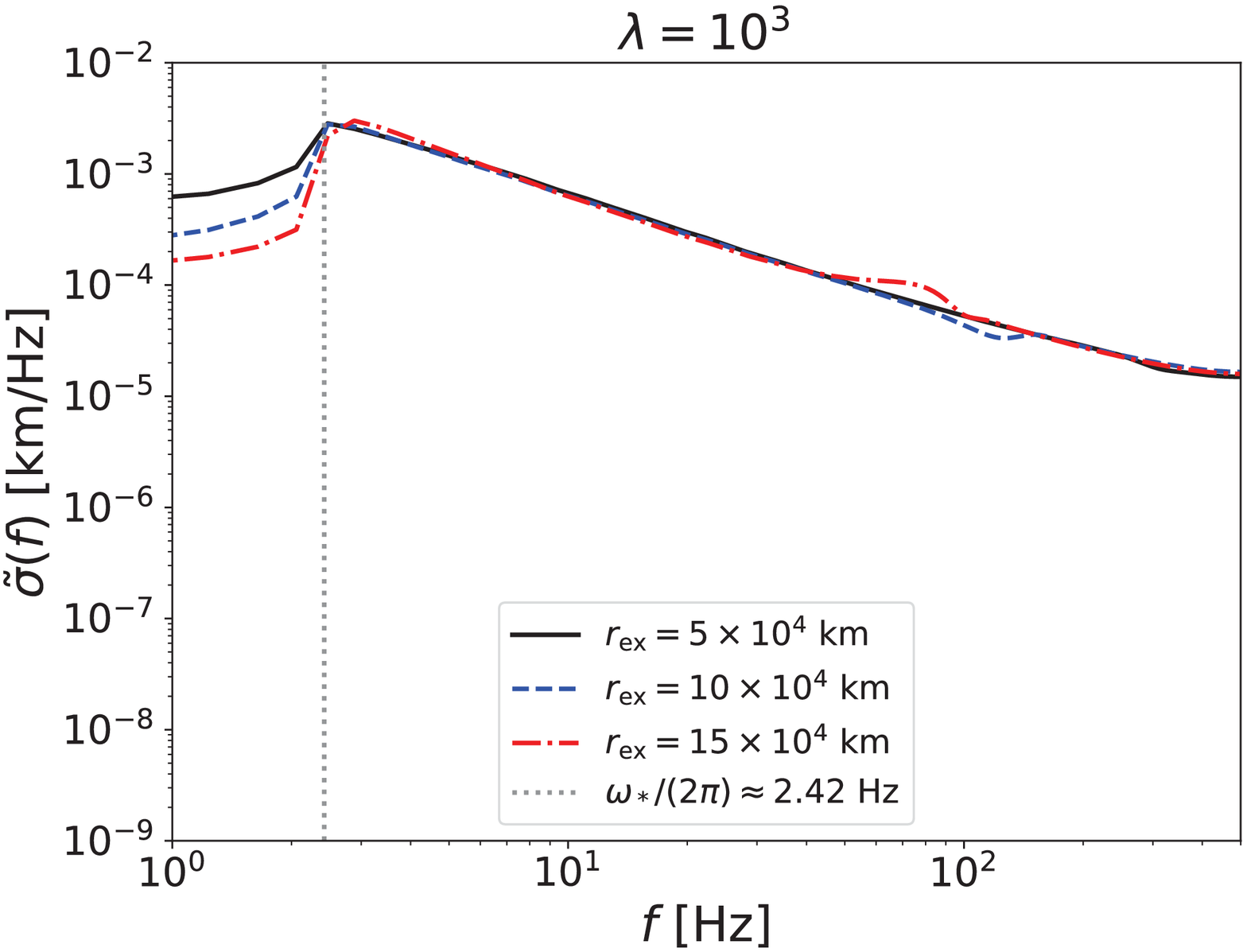}
	\caption{Scalar GW waveforms for WH12 extracted at different radii $r_{\rm ex} = (5,10,15)\times 10^4$~km. } \label{rext_NS}
\end{figure}	

In addition, the frequency-domain plot (the right panel of Fig.~\ref{benchmark_NS}) shows that the suppression is more significant for $f\lesssim f_* = 2.42$~Hz. Due to their smaller group velocities, the low-frequency modes spend more time propagating out of the (strong) interaction regime near the star so that they experience more suppressions by self-interactions. To illustrate this effect, we plot $\sigma$ at three radii $r_{\rm ex} = (5,10,15)\times 10^{4}$~km with $\lambda=10^3$ in Fig.~\ref{rext_NS}. These distances lie in the wave zone~\cite{Rosca-Mead:2020ehn} and are much larger than both the reduced Compton wavelength of the massive scalar $\lambda_C = c/(2\pi f_*)$ and the gravitational radius $r_G = GM_{\rm NS}/c^2$ with $M_{\rm NS}$ denoting the NS mass. In the left panel of Fig.~\ref{rext_NS}, we see that the time-domain signal becomes much more oscillatory since the dispersion during the propagation screens out the low frequency modes for a distant observer, as expected. When transformed into the frequency domain (the right panel of Fig.~\ref{rext_NS}), the modes with frequencies higher than $f_*$ remain unchanged, while the spectra below this critical frequency decay exponentially. This implies that only the high-frequency scalar signals can travel astrophysically long distance to be detectable on the Earth. Focusing on the high frequency signals, the choice of the extraction radius is irrelevant therefore.

\subsection{Numerical Results on Core Collapse to Black Holes}
Now we turn to simulation results for the progenitor model WH40. In order for the strong scalarization to occur in the proto-NS prior to the BH formation, we set the ST parameters as $\alpha_0 = 3\times 10^{-3}$ and $\beta_0 = -5$. We shall repeat the above analysis to see the effects of the self-interaction term in Eq.~(\ref{potential}) on the dynamics of the stellar evolution and the scalar-polarized GW radiations.

\begin{figure}[!htb]
	\centering
	\hspace{-5mm}
	\includegraphics[width=0.5\linewidth]{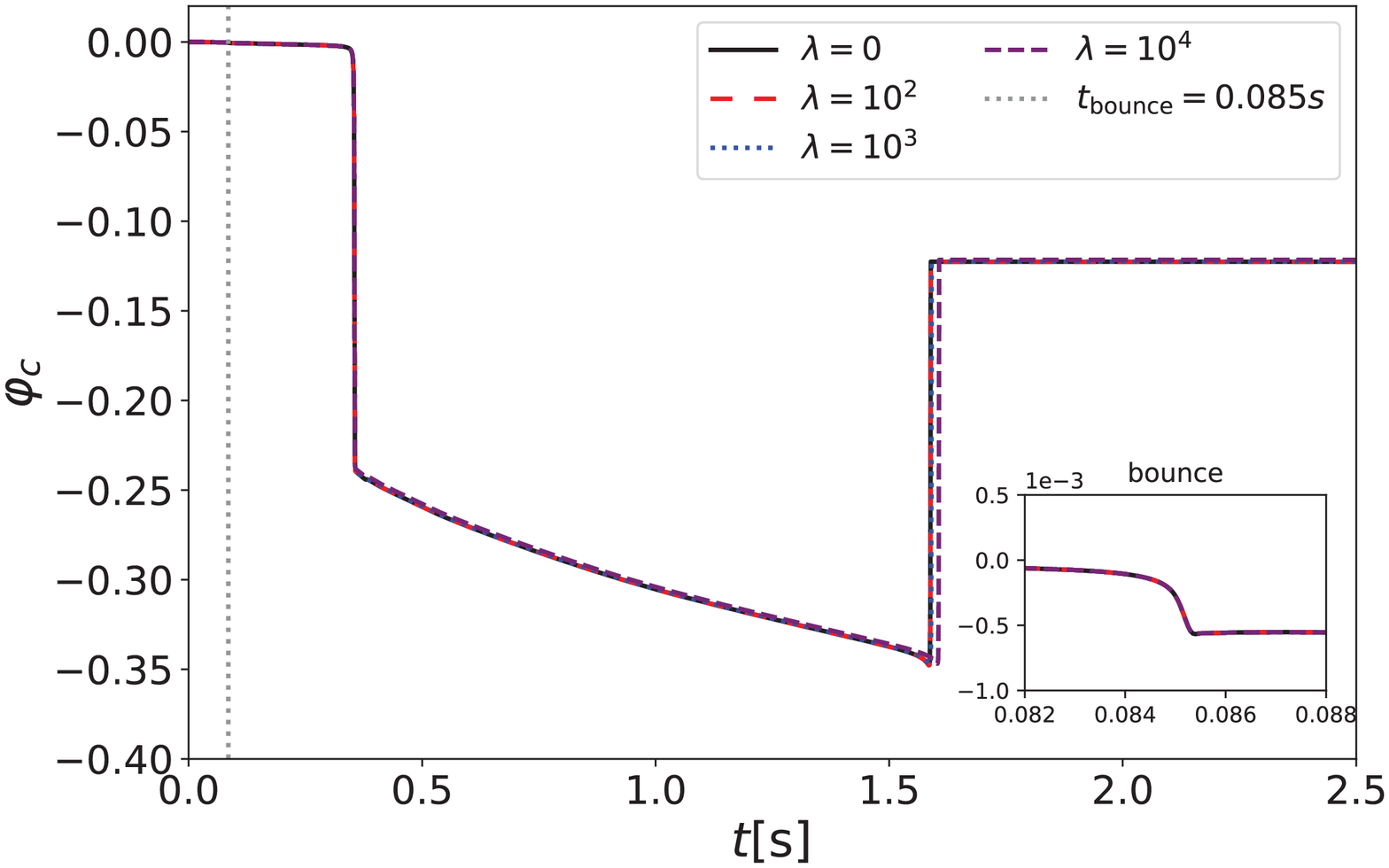}	
	\includegraphics[width=0.5\linewidth]{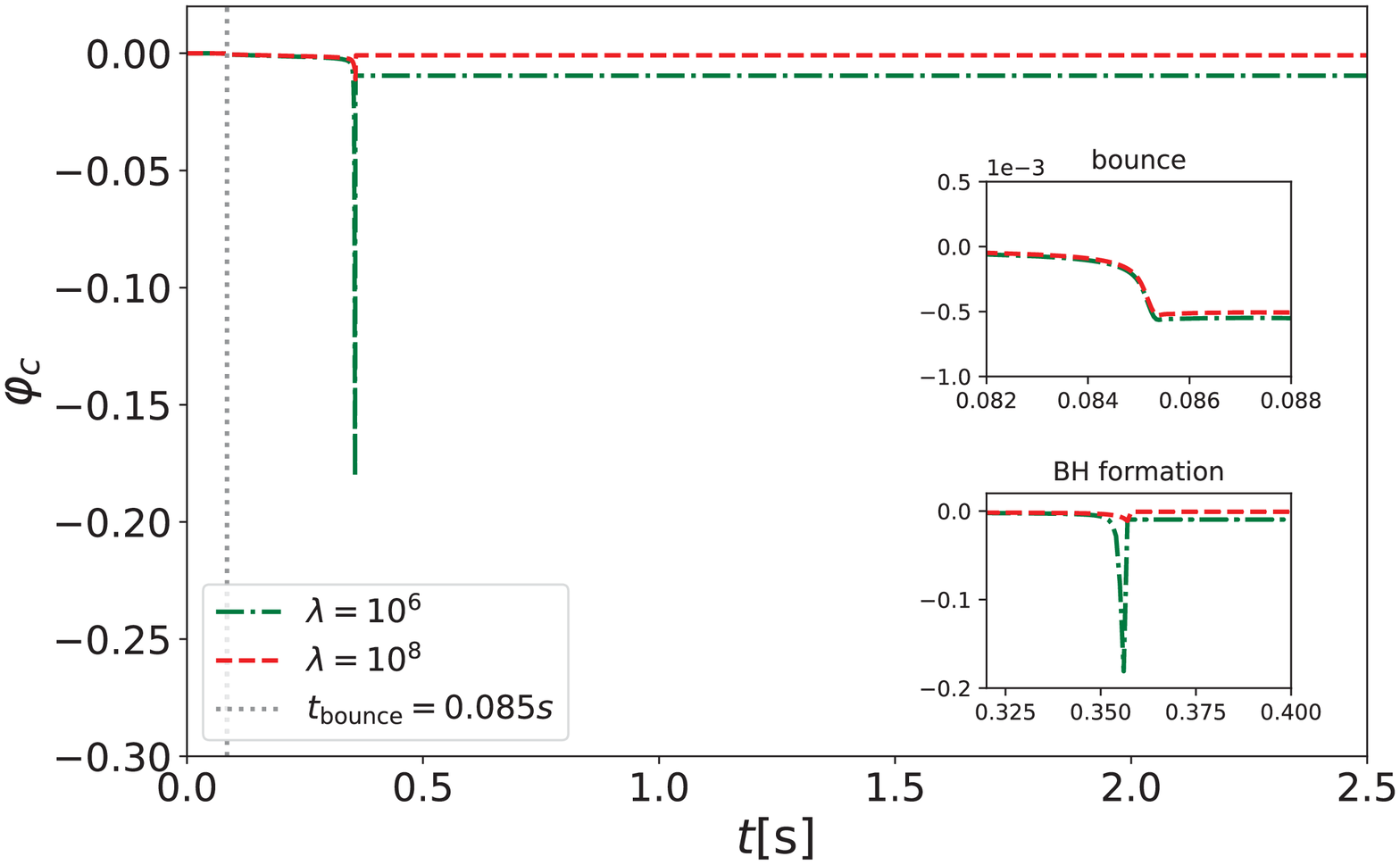}
	\caption{Evolutions of the central value of the scalar field $\varphi_c$ for the progenitor WH40 during the core-collapse into a BH for $\lambda = 0$, $10^2$, $10^3$ and  $10^4$ in the left panel and $\lambda=10^6$ and $10^8$ in the right panel, respectively, where the inset of left panel shows the scalar field behavior zoomed in on the core bounce part, while ones on the right panel correspond to the region near the core bounce and the BH formation, respectively.} \label{central_BH}
\end{figure}	

Fig.~\ref{central_BH} presents  the evolutions of the scalar field at the center of the compact objects for the WH40 progenitor. Obviously, the dynamics is much more complicated than the WH12 model. We begin our discussion with the non-self-interacting case, {\it i.e.}, $\lambda=0$. The stellar core bounce firstly produces a weakly scalarized NS with $\pc\sim -5\times10^{-4}$ at $t_{\text{w}} \simeq 0.085$~s, which then transits to a strongly scalarized NS at $t_{s}\simeq 0.35$ s due to the accretion of ambient materials. In this particular model, the continuous infalling material makes the stellar core massive enough to cause the gravitational instability, i.e., the mass of the remnant exceeds the maximal mass that can be supported by the EOS considered. Eventually, a BH forms at some point $t_{\text{BH}}$, which depends on $\lambda$ (see below).

The extent to which the scalarization is triggered and the lifetime of the strongly-scalarized state ($t_{\text{BH}}-t_{\text{s}}$) depend on $\lambda$. The left panel of Fig.~\ref{central_BH} shows the cases with a moderate level of $\lambda$, where we see that $\pc$ goes down to $\sim-0.25$ at $t_{\text{s}}$, and then the scalarizaion is progressively strengthened to $\pc\sim-0.35$ at the moment right before the BH formation. The strong scalarization lasts for $\sim1.24$ s, and is followed by a transient decrease of the scalarization at $t_{\text{BH}}$ in accordance with the BH no-hair theorem.  In addition, when $\lambda \lesssim 10^3$, we do not see any visible deviation from the case without self-interactions. As $\lambda$ increases to $\gtrsim10^4$, the degree of the scalarization during the strong-scalarization stage is weakened a little, but the duration $t_{\text{BH}}-t_{\text{s}}$ becomes a bit longer owing to a delay in the BH formation. Such trend continues and becomes more evident with increasing $\lambda$ until $\lambda\sim10^6$. In the right panel of Fig.~\ref{central_BH}, we plot the case with strong coupling strengths $\lambda=10^6$ and $10^8$. For $\lambda=10^6$, the degree of the scalarization is a bit slighter than the aforementioned case with the moderate $\lambda$ and the lifetime of the proto-NS becomes very short with $t_{\text{BH}}-t_{\text{s}}\lesssim0.01 \text{ s}$. When $\lambda= 10^8$, we find that the strongly-scalarized proto-NS stage completely disappears, leading to a scenario that a weakly-scalarized NS directly collapses to a BH.

\begin{figure}[!htb]
	\centering
	\hspace{-5mm}
	\includegraphics[width = 0.5 \linewidth]{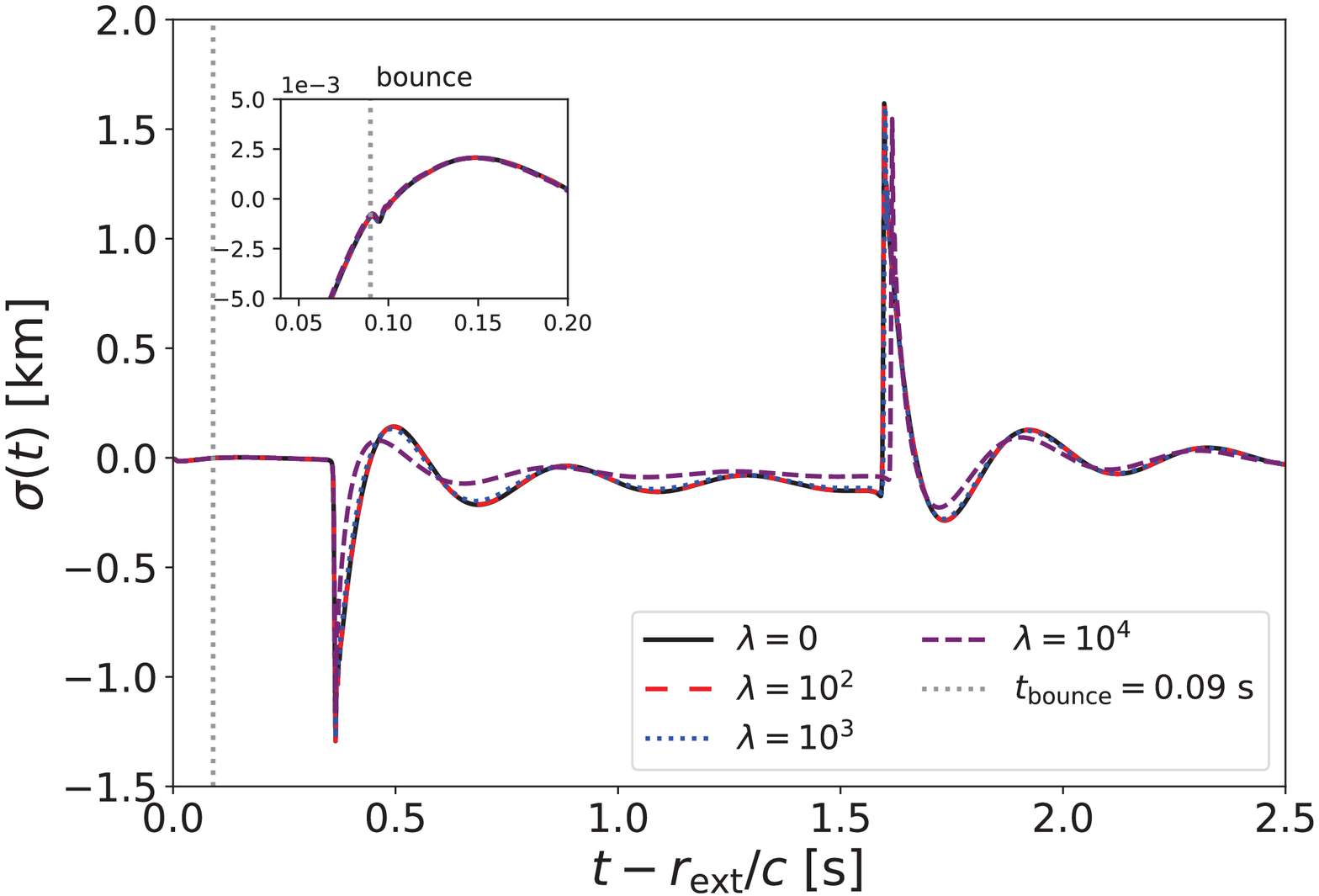}	
	\includegraphics[width=0.5\linewidth]{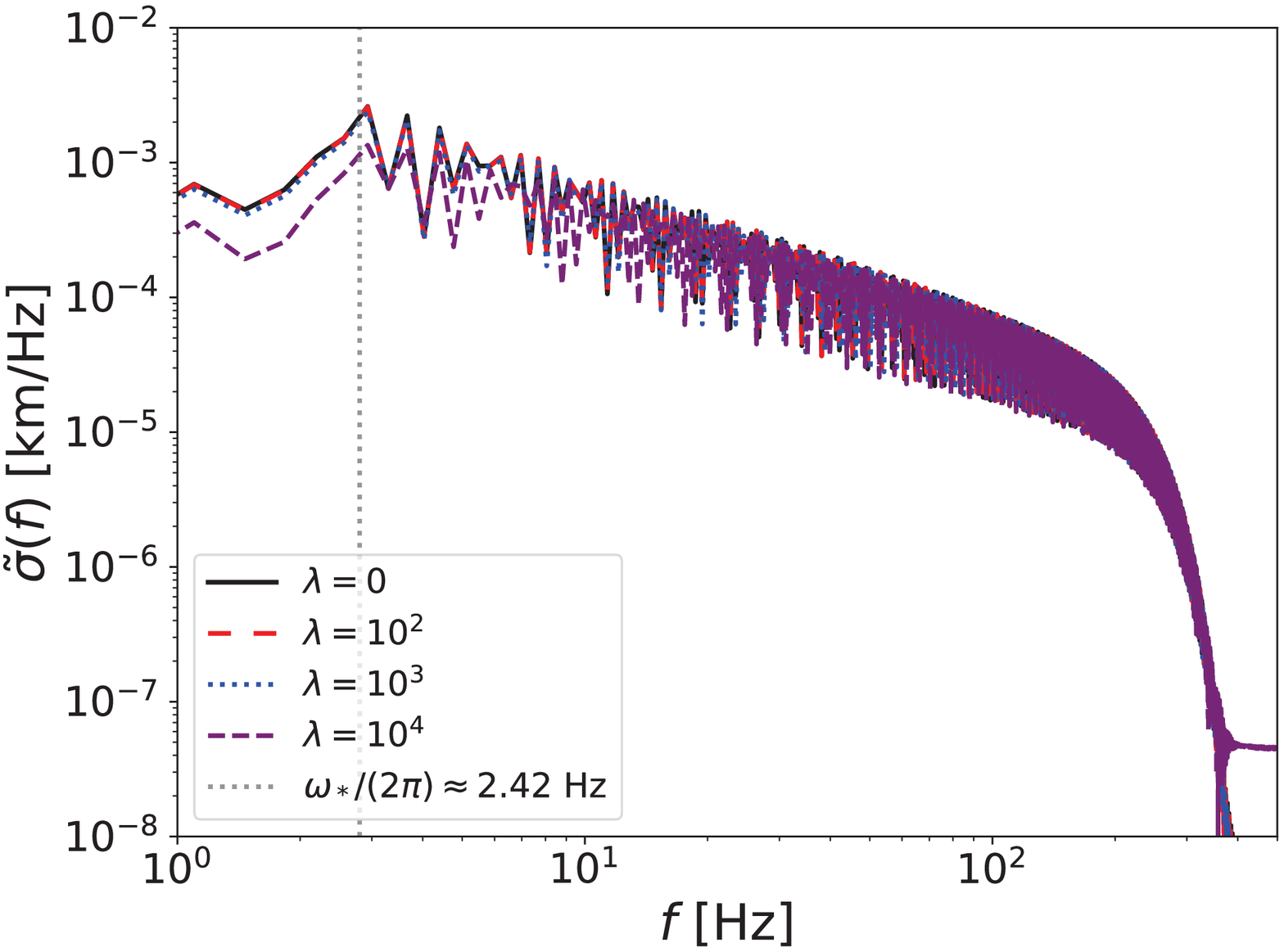}
	\caption{Scalar GW signals in the time domain (left panel) and frequency domain (right panel) for the progenitor model WH40 extracted at the radius $r_{\rm ex} = 5\times 10^{4}$~km for different values of the self-interaction coupling $\lambda =$0, $10^2$, $10^3$ and $10^4$, respectively. The inset in the left panel illustrates the scalar GWs around the core bounce which generates a weakly-scalarized NS. } \label{benchmark_BH}
\end{figure}	

\begin{figure}[!htb]
	\centering
	\hspace{-5mm}
	\includegraphics[width = 0.5 \linewidth]{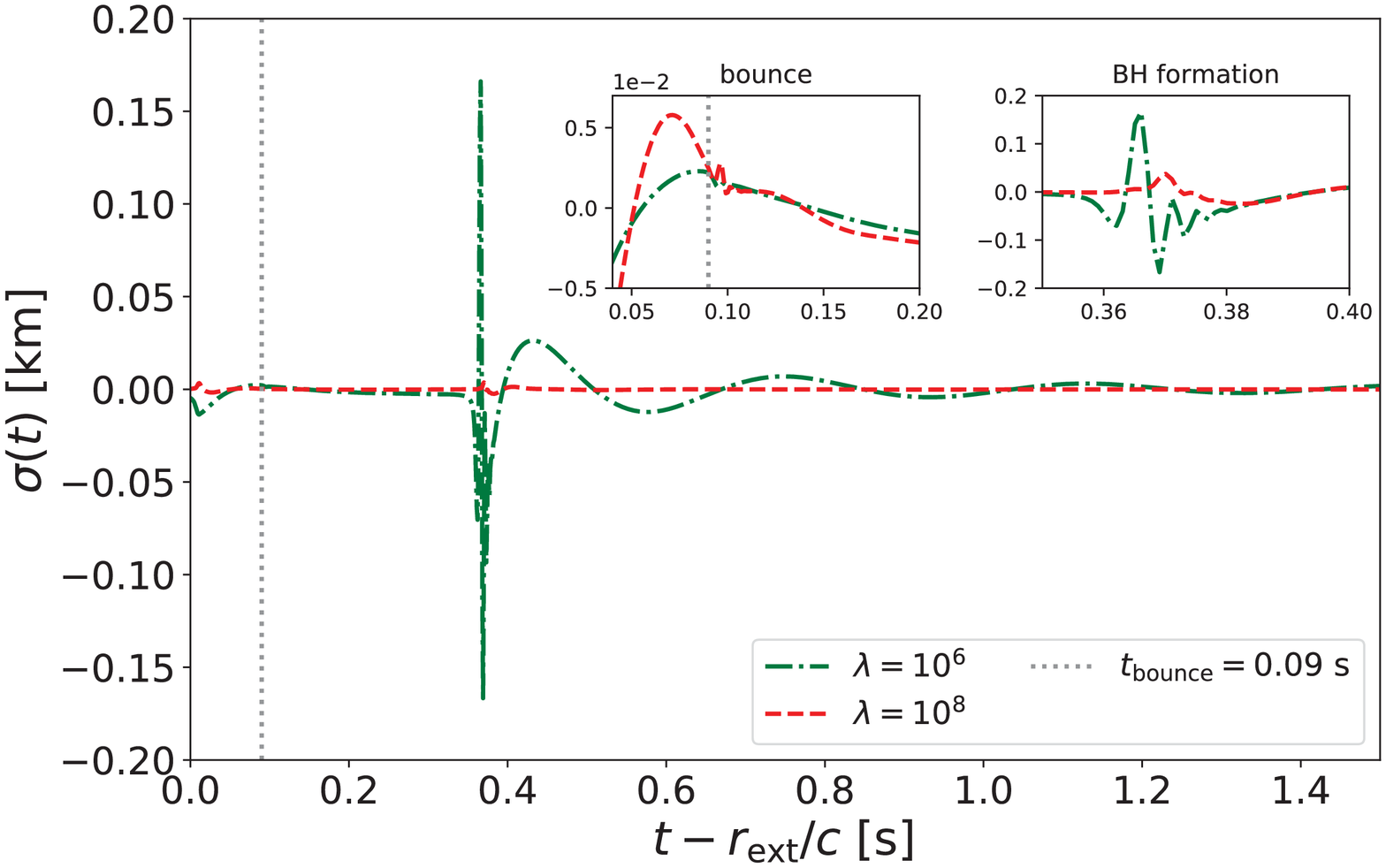}	
	\includegraphics[width=0.5\linewidth]{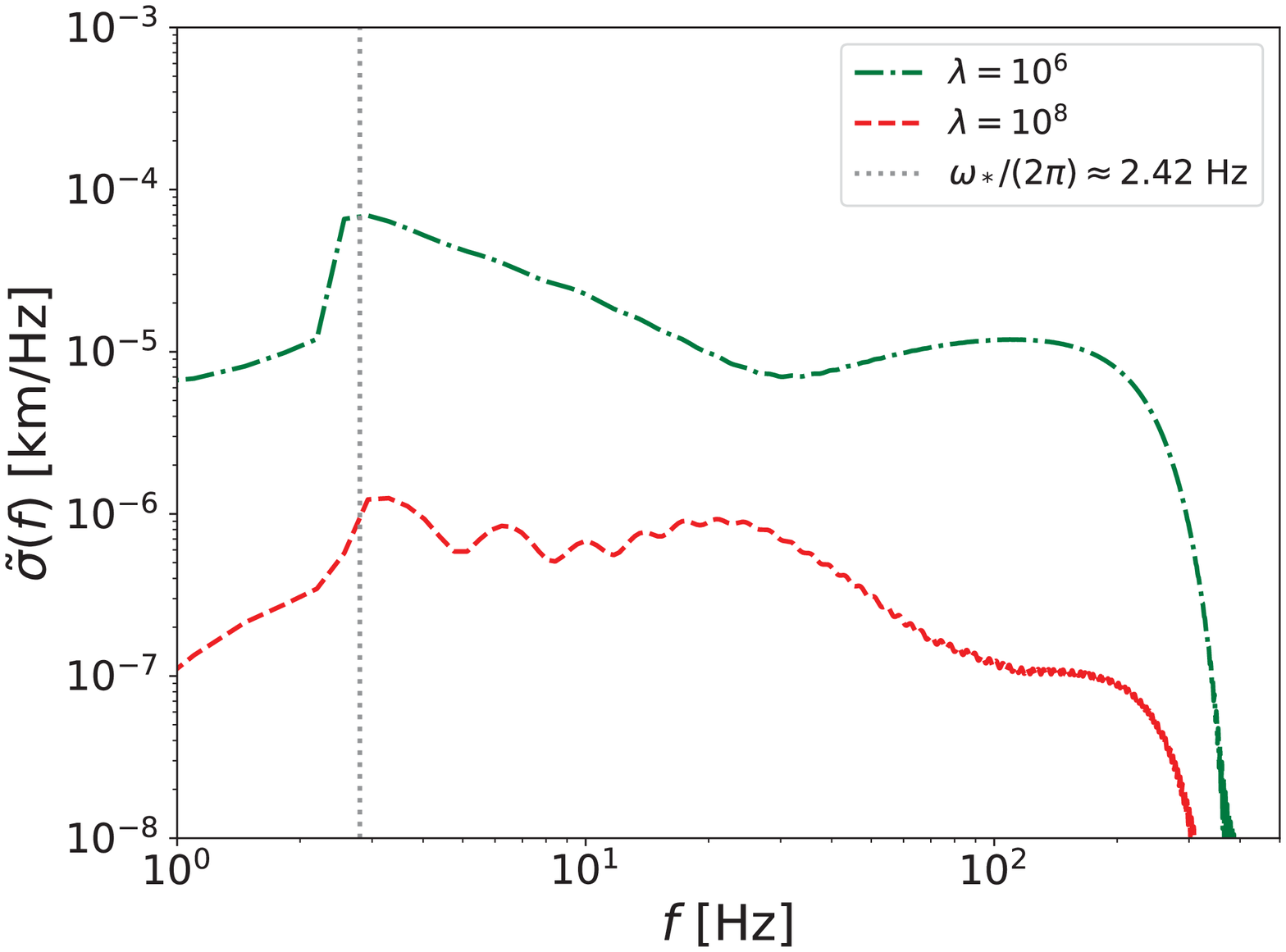}
	\caption{Label is the same as  Fig.~\ref{benchmark6_BH}, except for $\lambda=10^6$ and $10^8$. The two insets in the left panel correspond to the scalar GW signals at the core bounce and the BH formation, where we have artificially enlarged the field values by a factor of 10 for the case with $\lambda=10^8$ in order to make its dynamical evolution clearer. } \label{benchmark6_BH}
\end{figure}	

Having shown the influence of $\lambda$ on $\pc$, we now turn to investigate how $\lambda$ affects $\sigma$, especially $\sme$.
In Figs.~\ref{benchmark_BH} and \ref{benchmark6_BH}, we present $\sme$ in both the time and the frequency domains. As illustrated in the left panel of Fig.~\ref{benchmark_BH}, we can find clear time-domain signatures corresponding to the transitions between different stages of the multi-stage BH formation scenario. In the inset figure, we see a small variation in the amplitude at the retarded time $t_{r}\equiv t-r_{\rm ex}/c =0.085$~s, which matches the timing of the core bounce. The sudden drop at $t_r\simeq 0.35$~s and the sharp peak around $t_r \simeq 1.59$~s reflect the transition from weakly- to strongly-scalarized configurations and the descalarization caused by the BH formation. Moreover, compared with the non-self-interacting case, the evolution of $\sme$ keeps intact for $\lambda\lesssim 10^3$. When $\lambda$ increases to $10^4$, the suppression of the amplitude for GW signals can be seen in both the time and frequency domains. On the other hand, as shown in the left panel of Fig.~\ref{benchmark6_BH}, besides the small structure at the core bounce, we only see some rapid oscillations around the BH formation for $\lambda\gtrsim 10^6$, which indicates the disappearance of the strongly-scalarized NS stage. Furthermore, the right panel of Fig.~\ref{benchmark6_BH} illustrates that the scalarization is diminished greatly by several orders for $\lambda =10^6$ and $10^8$ in comparison with the case of $\lambda=0$.  


\section{Scalar GW Propagation and its Detectability}\label{GW_PSD}

\subsection{Scalar GW Signals}
In the ST gravity, the scalarization contributes the strain amplitude~\cite{Will:2014kxa}:
\begin{eqnarray}\label{eq:phistrain}
h_s = h_B-h_L\,,
\end{eqnarray}
to the GW signal of the angular frequency $\Omega$, where
\begin{eqnarray}
h_B = 2\alpha_0 \varphi\,,
\end{eqnarray}
is the so-called breathing mode which is transverse and scalar-polarized, and 
\begin{eqnarray}
h_L = \left(\frac{\omega_*}{\Omega}\right)^2 h_B = 2\alpha_0 \left(\frac{\omega_*}{\Omega}\right)^2 \varphi\,,
\end{eqnarray}
is the longitudinally polarized mode, respectively. For a GW interferometer, the detector responses to the scalarization-induced GW strain through
\begin{align}\label{eq:gwresponse}
	h(t) = F_\circ [\theta(t), \phi(t)] h_s(t),
\end{align}
where $F_\circ \equiv -\sin^2\theta \cos 2\phi/2$ is the interferometer antenna pattern~\cite{Will:2014kxa,Babusci:2001tw} which depends on the sky location $(\theta, \phi)$ of the source in the internal frame of the detector. Note that these two scalar polarizations share the identical antenna pattern function up to a sign~\cite{Will:2014kxa}, so that they cannot be distinguished experimentally. For simplicity, we take the sky averaged rms value of the antenna pattern function~\cite{Rosca-Mead:2020ehn}
\begin{eqnarray}
\bar{F}_\circ = \sqrt{\iint d\theta d\phi \sin\theta F^2(\theta,\phi)} = \sqrt{4\pi/15}
\end{eqnarray} 
as the canonical value. 

Currently, the weak-field experiments in our solar system have strongly constrained the scalar mass to be $\mu>{\cal O}(10^{-15})$~eV, corresponding to a lower frequency bound of ${\cal O}(0.1~{\rm Hz})$ via Eq.~\eqref{eq:comp}. The GW modes above this bound lies in the sensitive range of ground-based GW detectors, such as LIGO-Virgo~\cite{TheLIGOScientific:2014jea, TheVirgo:2014hva,Aasi:2013wya}, Cosmic Explorer (CE)~\cite{Evans:2016mbw} and Einstein Telescope (ET)~\cite{Punturo:2010zz}. In the literature, the detectability of the GW signal in Eq.~\eqref{eq:phistrain} for a single detector is characterized by the corresponding SNR which is defined as follows~\cite{Moore:2014lga}:
\begin{eqnarray}\label{eq:SNR}
\rho^2 = 4\int^\infty_0 df \frac{|\tilde{h}(f)|^2}{S_n(f)}\,.
\end{eqnarray}
where $\tilde{h}(f)$ denotes the Fourier transformed function of the response in Eq.~\eqref{eq:gwresponse}, while $S_n(f)$ is the one-sided noise power spectral density of the instrument given by the corresponding experimental collaboration. 

\subsection{Scalar GW Propagation}
After the scalar GWs [Eq.~\eqref{eq:phistrain}] are generated from the stellar core-collapse in our Galaxy, they propagate over an astronomically long distance of ${\cal O}(10~{\rm kpc})$ before they are detected on the Earth. According to the action in Eq.~(\ref{ActionEinstein}), the propagation of the scalar GW far away from its source can be well approximated by the wave equation in the nearly flat space:
\begin{eqnarray}
\left\{\begin{array}{c}
\frac{\partial^2 \sigma}{\partial t^2} - \frac{\partial^2 \sigma}{\partial r^2} + \frac{\mu^2}{\hbar^2} \sigma + \frac{3\lambda \mu^2}{2\hbar^2}\frac{\sigma^2}{r} =0\,, \quad {\rm as}\quad \sigma\geq 0\,,\\
\frac{\partial^2 \sigma}{\partial t^2} - \frac{\partial^2 \sigma}{\partial r^2} + \frac{\mu^2}{\hbar^2} \sigma - \frac{3\lambda \mu^2}{2\hbar^2}\frac{\sigma^2}{r} =0\,, \quad {\rm as}\quad \sigma < 0\,,
\end{array}\right.
\end{eqnarray}
where the equation is written in the spherically symmetric coordinate with $\sigma\equiv r\varphi$. The interaction term drops faster than the mass term by one power of $r$, and the mass term will dominate after a certain distance. The subsequent long-distance propagation is thus barely affected by the self-interactions. In this case, the description of the scalar GWs can be further reduced to the following wave equation without interactions
\begin{eqnarray}
\frac{\partial^2 \sigma}{\partial t^2} - \frac{\partial^2 \sigma}{\partial r^2} + \frac{\mu^2}{\hbar^2} \sigma =0\,,\quad {\rm as} \quad r \to \infty\,.
\end{eqnarray} 
  
As pointed out in Refs.~\cite{Sperhake:2017itk,Rosca-Mead:2020ehn}, due to the presence of the mass term, the GW signal becomes more and more oscillatory with time. The dispersive nature of the propagation introduces tremendous difficulties in the numerical calculations of the GW evolution in space. In the present work, we adopt the stationary phase approximation (SPA)~\cite{BenderMath} to evaluate the scalar GW waveform at large radii. The scalar GW signal at $r_{\text{ex}}$ can be expanded by the Fourier series, in which the component with frequency $\omega$ reads
\begin{eqnarray}
\tilde{\sigma}(\omega; r_{\rm ex}) = {\cal A}(\omega; r_{\rm ex}) e^{i\Psi(\omega)}\,,
\end{eqnarray}    
where ${\cal A}(\omega;r_{\rm ex})$ and $\Psi(\omega)$ are the scalar wave amplitude and phase, respectively. In the limit of $r\gg r_{\rm ex}$, the scalar GW signal is highly dispersive and becomes quasi-monochromatic at the frequency
\begin{eqnarray}\label{InvChirp}
\Omega(t) = \frac{\omega_* t}{\sqrt{t^2 - (r-r_{\rm ex})^2}}\,,\quad {\rm for} \,\, t> r-r_{\rm ex}\,.
\end{eqnarray}
Note that this frequency shows the signature of the inverse chirp~\cite{Sperhake:2017itk}, which means that high-frequency modes arrive earlier than low-frequency (but still higher than $\omega_*$) ones. Such a feature is originated from the fact that scalar GW modes with higher frequencies have larger group velocities, and thus reach us with less time. Furthermore, the SPA method also gives the following amplitude of the scalar GW at a large radii $r$~\cite{Sperhake:2017itk,Rosca-Mead:2020ehn}
\begin{eqnarray}
A(\Omega;r) = \sqrt{\frac{2[\Omega^2 - \omega_*^2]^{3/2}}{\pi\omega_*^2 (r-r_{\rm ex})}}{\cal A}(\Omega; r_{\rm ex})\,, 
\end{eqnarray} 
in which the frequency $\Omega$ is a function of time $t$ given by Eq.~(\ref{InvChirp}).

As argued in Ref.~\cite{Rosca-Mead:2020ehn}, for a (quasi-)monochromatic scalar GW signal, the SNR Eq.~\eqref{eq:SNR} can be estimated as follows
\begin{eqnarray}
\rho \approx \sqrt{\frac{S_o(f)}{S_n(f)}} \,.
\end{eqnarray}
Here, the spectral power density of the GW signal is given by
\begin{eqnarray}\label{eq:specpwdens}
\sqrt{S_o(f)} =  {\sqrt{T} \alpha_0 \bar{F}_\circ}\left(\frac{A(2\pi f; D)}{D}\right) \left[1-\left(\frac{\omega_*}{\Omega}\right)^2\right]\,,
\end{eqnarray}
where $D$ is the distance between the star and the Earth, while $T$ denotes the duration of the observation.

\subsection{Scalar GW Detectability}
For the scalar GW signals in the ST gravity, there exist several search strategies~\cite{Sperhake:2017itk}, such as monochromatic continuous-wave searches, stochastic searches, and burst searches. For instance, for a CCSNe that is detected by optical observations or all-sky GW searches, the best strategy is through the continuous quasi-monochromatic GW search~\cite{Riles:2017evm} pointed to the location of that source. In this subsection, we present our prediction of the detectabilities of such source-targeted continuous searches of the scalar GW signals for the progenitor models WH12 and WH40. 

\begin{figure}[!htb]
	\centering
	\hspace{-5mm}
	\includegraphics[scale=0.4]{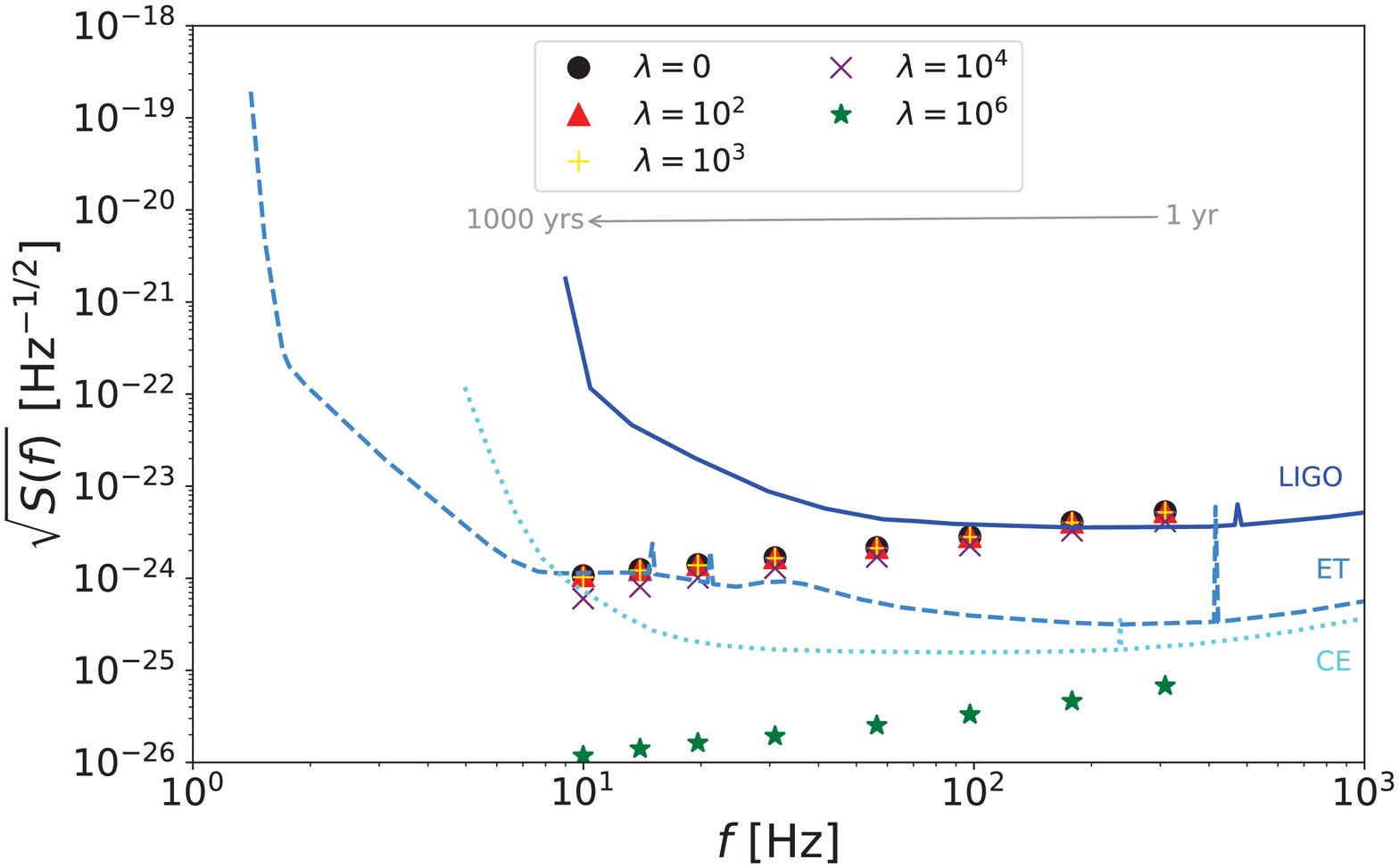}	
	\includegraphics[scale=0.4]{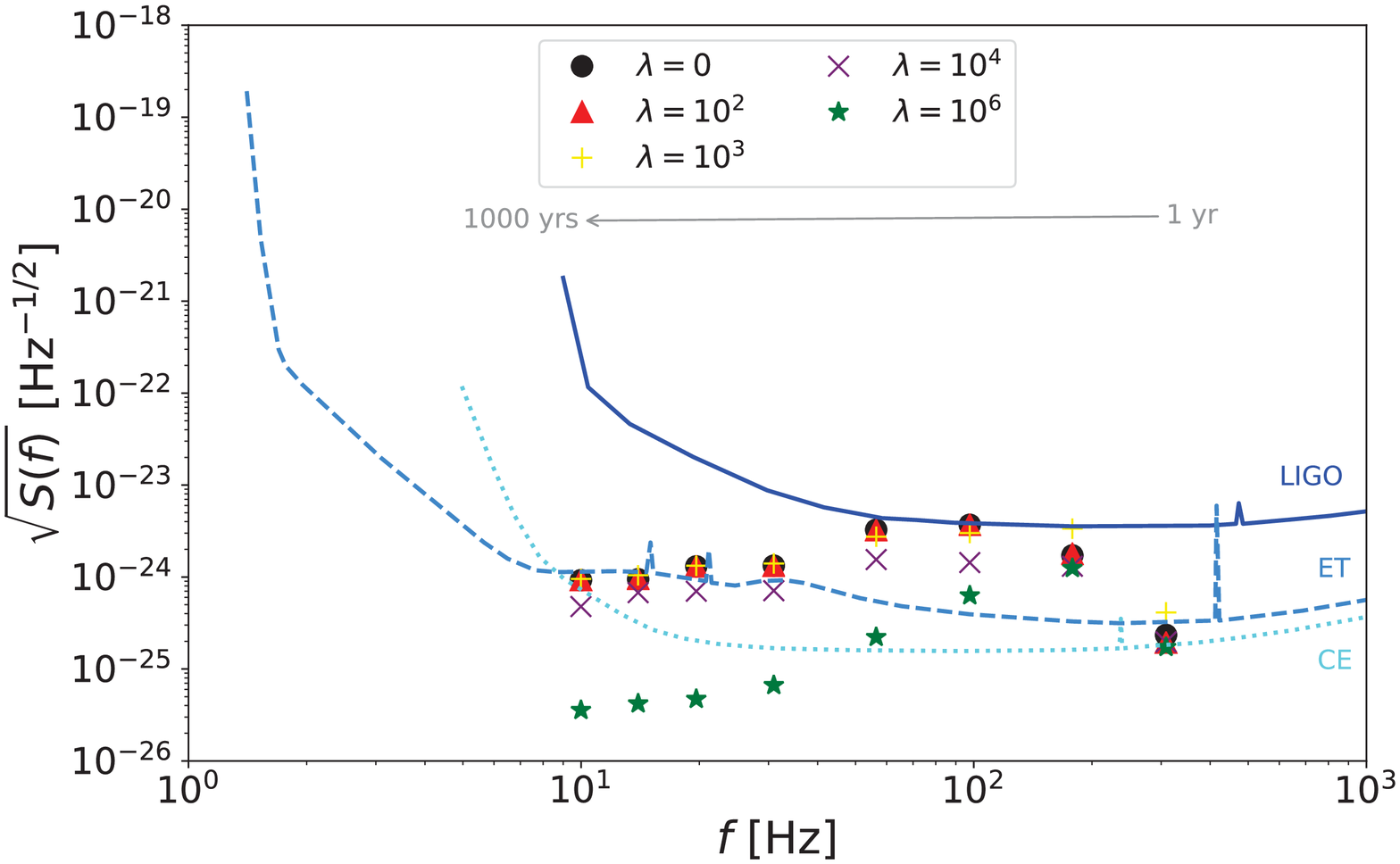}	
	\caption{Power spectrum densities of the core collapse into a NS for the progenitor models WH12 (upper panel) and into a BH for WH40 (lower panel) with  the self-interaction couplings $\lambda =$0, $10^2$, $10^3$, $10^4$ and $10^6$, respectively, where the points of various types correspond to the inverse-chirp scalar GW signals observed at different times after the supernova: $t=$1, 3, 10, 30, 100, 250, 500, and 1000 years from right to left on each plot. In comparison, we also plot the noise power spectral density of LIGO, ET, and CE.} \label{spectral}
\end{figure}	

By employing the scalar GW waveforms extracted at $r_{\rm ex} = 5\times 10^4$~km to Eq.~\eqref{eq:specpwdens}, we can obtain the spectral power densities for both progenitor models. In Fig.~\ref{spectral}, we plot $\sqrt{S_o(f)}$ at several time points after the GW emission with various triple-scalar coupling strengths. 
Here, the distance between the Earth and the source is assumed to be of the galactic scale, {\it i.e.}, $D=10$~kpc, and the observational duration is $T$= 2 months. For both WH12 and WH40, we can see that (i) the quasi-monochromatic signal slowly moves to low frequencies in accord with the inverse-chirp formula in Eq.~(\ref{InvChirp}), (ii) $\sqrt{S_o(f)}$ varies within two orders over time for a given coupling strength, and (iii) the triple-scalar self-interaction suppresses the magnitude of the scalar gravitational radiation generated by the stellar core collapses. The predicted scalar GW signals are similar to the non-interacting case for up to $\lambda \sim 10^3$, whereas we observe a small reduction in the signal amplitude when $\lambda$ increases to $\sim10^4$. When $\lambda \gtrsim 10^6$, the scalar-polarized GW is suppressed by two orders for the WH12 model, while for the WH40 progenitor the suppression is frequency-dependent, viz.~the low-frequency modes decrease in amplitude more than the high-frequency part of the spectrum.    

On top of the scalar GW strain of our models, Fig.~\ref{spectral} also shows the noise curves of the current LIGO detectors, and the near-future ET and CE experiments, whereby we can estimate the SNRs of signals. It turns out that, even though the scalar GW signal amplitude of the WH12 model with $\lambda \lesssim 10^4$ observed for 2 months in the first 3 years after the core collapse can be higher than the LIGO noise curve, the SNR in this optimal case can only reach $\rho \simeq 2$ at most, which is unlikely to be detected. For other choices of the coupling constant values, progenitor models and detection times would result in power spectral densities $\sqrt{S_o}$ lower than the LIGO sensitivity curve, implying that it is impossible to measure them with LIGO. In contrast, with the future ET and CE detectors, some signals may reach SNRs of $\rho \sim {\cal O}(100)$ and remains visible for several centuries.      


\section{Conclusions and Discussions}\label{Conclusion}
The stellar core collapse provides us with valuable tests of ST theories due to the possible formation of the NS in the intermediate stage or as the final state of the process. The spontaneous scalarization may be induced in those NS and produces strong scalar or monopole GW signals. In the present work, we have studied the effects of the triple-scalar self-interaction defined in Eq.~(\ref{potential}) on the scalar field dynamics and the subsequent scalar GW generation. In order to achieve this goal, we have performed simulations based on the open-source code {\tt GR1D} which is extended to the massive ST gravity with the triple-scalar self-interaction. We have focused on two specific pre-SN models with the primordial metallicity in the WH list, viz.~WH12 and WH40, corresponding to stars with ZAMS mass of $12M_\odot$ and $40M_\odot$, respectively. These two progenitors are representative of two typical collapse processes; WH12 collapses into a NS, while WH40 ends up to a BH. The ST parameters $\alpha_0$ and $\beta_0$ are chosen specifically for each model so that the strong scalarization can take place in simulations. 

As a result, we have found that the triple-scalar self-interaction can generically suppress the degree of the scalarization during the collapse, which is consistent with the conclusions in Refs.~\cite{Cheong:2018gzn,Rosca-Mead:2019seq}. Such suppressions can be shown either from the scalar dynamics at the center of the star or from the produced scalar GW signals. Furthermore, the suppression is considerable only when the self-interaction coupling $\lambda$ becomes sufficiently large, {\it i.e.}, $\lambda \gtrsim 10^3$ for WH12 and $\lambda \gtrsim 10^4$ for WH40. More interestingly, the self-interaction can even alter the scalar evolutionary history of the collapse for the WH40 model by completely eliminating the strongly scalarized stage during the BH formation as $\lambda$ reaches a critical value of ${\cal O}(10^6)$. 

For the massive scalar GWs propagate over astrophysically long distances, the self-interaction effects have been shown to be negligible, and the dispersion of GWs is solely determined by the scalar mass. It turns out that scalar GW signals detected on the Earth have the inverse-chirp feature as that have been suggested by massive ST theories without self-interactions \cite{Sperhake:2017itk,Geng:2020slq}. We have accessed the detectability of the scalar GW signals produced by the galactic WH12 and WH40 pre-SN sources by estimating their SNRs. Although it is unlikely to observe the inverse-chirp signals for both progenitors with the current LIGO detectors, signals of these two models can reach SNRs of ${\cal O}(100)$ with the future ET and CE detectors and remain visible for several centuries. 

In this study, our simulations have been performed with fixed nuclear matter EOS parameters and specific values of the ST gravity parameters $(\alpha_0,\beta_0)$. However, it is well known that the nuclear matter property would significantly affect the hydrodynamic evolution of matter fields during the core collapse \cite{Dimmelmeier:2007ui, Dimmelmeier:2008iq}, which would in turn dramatically modify the dynamics of the scalar field and the generated GW waveforms. Additionally, it has been pointed out in Ref.~\cite{Rosca-Mead:2020ehn} that the stellar collapse scenarios changes with the different choices of the ST parameters ($\alpha_0$, $\beta_0$). Especially, as the value of $\beta_0$ becomes more and more negative, the degree of scalarization tends to be intensified and the corresponding GW signals are expected to be stronger. Nevertheless, the full exploration of all these aspects is well beyond the scope of the present article, and will be deferred to the future work.

\section*{Acknowledgments}
This work is supported in part by 
the National Key Research and Development Program of China (Grant No. 2020YFC2201501).
DH is supported by National Natural Science Foundation of China (NSFC) under grant No. 12005254 and the Key Research Program of the Chinese Academy of Sciences under grant No. XDPB15. H.-J. K appreciates the financial support of the Sandwich grant (JYP) No.~109-2927-I-007-503 by DAAD and MOST.


\end{document}